# Research on the Dynamic Evolution and Influencing Factors of Energy Resilience in China

Tie Wei, Youqi Chen, Zhicheng Duan

**Abstract:** Energy security is the guarantee for achieving the goal of carbon peaking and carbon neutrality, and exploring energy resilience is one of the important ways to promote energy security transition and adapt to changes in international and domestic energy markets. This paper applies the combined dynamic evaluation method to measure China's energy resilience level from 2004-2021, analyses the spatio-temporal dynamic evolution of China's energy resilience through the center of gravity-standard deviation ellipse and kernel density estimation, and employs geo-detectors to detect the main influencing factors and interactions of China's energy resilience. The study finds that: (1) China's energy resilience level generally shows a zigzagging forward development trend, and the spatial imbalance characteristic of China's energy resilience is more obvious. (2) The spatial dynamics of China's energy resilience level evolves in a northeast-southwest direction, and the whole moves towards the southwest, with an overall counterclockwise trend of constant offset. (3) When the energy resilience level of neighboring provinces is too low or too high, it has little effect on the improvement of the energy resilience level of the province; when the energy resilience level of neighboring provinces is 1-1.4, it has a positive spatial correlation with the energy resilience level of the province, and the synergistic development of the provinces can improve the energy resilience level together; the time span does not have a significant role in promoting the improvement of the energy resilience level among neighboring provinces. (4) GDP, the number of employees, the number of employees enrolled in basic pension and medical insurance, and the number of patent applications in high-tech industries have a more significant impact on China's energy resilience, while China's energy resilience is affected by the interaction of multiple factors.

# 1 introduction

In the post-epidemic era, the superposition of various factors such as scientific and technological progress, economic transformation and geopolitical conflicts has led to significant changes in the global energy supply and demand pattern and security environment, and all countries in the world, especially developing countries, are generally facing an unprecedentedly serious energy security situation [1]. Although China's energy supply has developed steadily in recent years, energy security problems are still serious. The energy development is unbalanced and insufficient, the new energy development technology has fallen into a bottleneck, fossil energy further clean and low-carbon use is difficult, the energy system mechanism is not perfect, the global energy pattern of increasing variables[2], some energy varieties of foreign dependence[3] and other energy security issues are not conducive to the realization of China's carbon peak carbon neutral goals, energy security is of vital importance to China's current modernization and development[4].

The traditional mechanism of energy security focuses on the problem of energy supply connectivity, which is obviously difficult to adapt to these new changes. Therefore, a new logic of energy security theory is needed to meet the new challenges of energy security[5]. The new concept of energy security, according to which the transition to energy security is the fundamental choice to solve the problem of energy poverty, is a major change in the components of the energy system. The most prominent manifestation is the adaptation and optimization of the energy structure and the advanced way of energy utilization, a process that is directly related to the scale, cost, security and reliability of the development, transmission and use of new energy, as well as the degree of clean and efficient use of fossil energy sources[6]. It can be said that the transition to energy security is an important way to cope with energy security problems such as the continuous growth of energy demand, the more traditional structure of energy production and the high proportion of high energy consumption industries[7]. However, the transition to energy security is a long-term process. In the face of political, market, emergency and other shocks, the energy system must have sufficient resilience to allow for a flexible response and rapid recovery in advance. For this reason, the United Nations energy resilience into the "2030 Agenda" to address the key issues[8], that the improvement of energy resilience is the world's countries to achieve energy security transition, security and carbon reduction of the road, but also to adapt to the changes in the international energy market is an important way[9].

Therefore, in the context of the new development of energy security, the transformation of China's energy security has become an inevitable trend. In this process, how to improve the resilience of energy supply has become a key link. In order to improve energy resilience, it is necessary to scientifically study the development characteristics and trends of China's energy resilience, and conduct in-depth research on the influencing factors of energy resilience, so as to formulate policies to improve energy resilience in a more scientific and targeted manner, thus promoting the transformation of China's energy security.

## 2 Literature review

The concept of resilience has been defined in many ways in previous studies, but there is no consensus[10], and its definition and interpretation varies from discipline to discipline[11,12]. Resilience is originally a direct translation of the Latin word "resiliere"[13], a concept first used in physics and mathematics[14], medicine and psychology[15], as a measure of stability indicating the ability of an object to survive a shock or trauma and return to a state of equilibrium in time[10,16], was later introduced into ecology to emphasize the importance of a system's ability to survive shocks by absorbing disturbances without losing the pre-disturbance relationships that govern the components of the system[16,17,18], and eventually appeared in economics and sociology[19], and was subsequently introduced into a wide range of research fields[20].

While The concept of energy resilience is interpreted in different ways, it is generally considered to refer primarily to the ability of an energy system to maintain a normal level of performance[20,21], the ability to sustain, cope with and overcome disruptions caused by economic, social, environmental and institutional shocks[22], and the adaptive capacity to continuously improve through learning and adapting to change[23].

Research on energy resilience currently focuses on the study of theories and systems rather than empirical analysis, including the development of frameworks for conceptualizing energy resilience, the assessment of values and attitudes towards changing energy systems[20], the measurement of assessment systems[24], assessment scales[25,26] and indices[27], the evaluation of energy resilience strategies[28] and case studies on energy resilience[29,30], etc. Theories and systems can provide a basis for empirical analysis, and empirical research on energy resilience is a future research trend in related fields.

In terms of measurement, energy resilience has similarities with economic[31,32], ecological[33,34], urban[35,36], electricity[37,38], transportation[39,40], and psychological[41,42] resilience. Many studies have measured energy resilience by constructing a system of indicators, and the existing energy resilience indicators mainly focus on the characteristics of energy itself, such as robustness, speed and redundancy[43], energy access index, renewable energy index and energy efficiency index[27]. Some researchers have also selected indicators at the risk resilience level of energy, such as adaptability, absorption and resilience[44,45]. On this basis, Toroghi and Valerie constructed a 5R system of energy resilience indicators, which includes robustness, redundancy, resourcefulness, speed and adaptability[46]. In addition, some researchers have also measured resilience from the perspective of service quality[47] and cost factors[48].

Previous relevant studies have provided valuable experience and foundations for the development of this study in terms of ideas and methods. This study acknowledges and draws on the relevant approaches mentioned above, but also recognizes the limitations of current academic research on energy resilience issues. Firstly, there is no academic consensus on the definition and assessment system of energy resilience, and there are still relatively few studies on energy resilience, most of which focus on theories and frameworks, lacking further analysis and empirical studies. Secondly, energy resilience needs to be researched from the two dimensions of time and space, but the current research perspective is heavier, which is not conducive to the development of energy resilience trends and development trends of in-depth research. Finally, the issue of energy resilience is related to the security and development of a region and a country, and it is very necessary to explore its influencing factors and take corresponding measures, but the influencing factors of energy resilience are less mentioned in the relevant research to date.

Compared to previous studies, this paper makes the following marginal contributions: First, this paper considers the dynamic evolution of China's energy resilience and explores the development characteristics and development trend of China's energy resilience in different spaces and times in an in-depth way, which provides ideas for future related studies; second, this paper considers the impact of the interaction of various factors on energy resilience and reflects the influencing factors of China's energy resilience in a more comprehensive way, which provides a decision-making basis for energy policy makers.

# 3 Methods and data sources

## 3.1 Construction of the indicator system

The construction of the indicator system in this paper follows the following principles: first, the principle of comprehensiveness, the selected indicators should reflect the characteristics of energy resilience as comprehensively as possible; second, the principle of representativeness, because it is impossible to select all relevant indicators in the process of constructing the system, so it is mainly to select the representative indicators; third, the principle of comparability, the main proportion of indicators and intensity indicators, and a small number of indicators of the size of the selection; fourth, the principle of operability, the construction of indicator system should have the effectiveness and applicability, taking into account the availability of data. The construction of indicator system should have the effectiveness and applicability, taking into account the availability of data, for the indicators with unavailable or missing data, this paper does not consider them for the time being[49].

In this paper, the process in which energy is confronted with disturbances is divided into three periods: before, during and after the disturbance. Energy resilience is defined as the ability of the energy system to anticipate, respond flexibly, adapt quickly to recovery, and continuously learn and optimize. Therefore, this paper constructs a comprehensive energy resilience assessment system based on four dimensions, namely energy supply level, research and development(R&D) level, transportation and informatization level, and ecological environment level, to measure the level of energy resilience of each province in China as shown in Table 1. The specific indicators are structured as follows:

(1) Energy supply level: The most important characteristics of energy resilience are reflected in energy productivity and sustainability[9], which is why energy intensity[50], the elasticity coefficient of energy production[51] and investments in the energy industry [52] are selected as indicators in this paper.

(2) R&D level: Energy resilience requires the improvement of energy use efficiency. The improvement of energy use efficiency mainly depends on technological progress, which largely depends on the R&D level of a country[53]. Therefore, indicators such as R&D intensity, R&D full-time equivalent and the level of the technology market development are selected in this paper.

(3) Transportation and informatization level: Both transportation and informatization level have a negative impact on energy intensity[54,55], at the same time, transportation and informatization level is the basis for measuring whether the energy system can anticipate and respond flexibly to disruptions. Therefore, the indicators selected in this paper are rail kilometers, freight volume and cell phone penetration.

(4) Ecological environment level: Resilience and ecological environment are closely related[22], and a good ecological environment can, to a certain extent, improve the ability of energy to withstand shocks and recover quickly. Therefore, this paper selects indicators such as carbon emissions intensity, forest cover and completion of industrial pollution abatement investments[51].

Tab.1 Comprehensive evaluation system of China's energy resilience

| Objective level | Primary indicators | Secondary indicators | Attributes | Weights |
| --- | --- | --- | --- | --- |
| Energy Resilience Synthesis Evaluation System | Energy supply Level | Energy intensity ($x_1$) | − | 0.084 |
| | | Energy production elasticity coefficient ($x_2$) | − | 0.002 |
| | | Energy industrial investment ($x_3$) | + | 0.126 |
| | R&D Level | R&D intensity ($x_4$) | + | 0.092 |
| | | R&D full-time equivalent ($x_5$) | + | 0.185 |
| | | Technology market development level ($x_6$) | + | 0.055 |
| | Transportation and informatization level | Rail kilometers ($x_7$) | + | 0.041 |
| | | Freight volume ($x_8$) | + | 0.212 |
| | | Cell phone penetration ($x_9$) | + | 0.010 |
| | Ecological environment level | Forest cover ($x_{10}$) | + | 0.124 |
| | | Carbon emission intensity ($x_{11}$) | − | 0.030 |
| | | Industrial pollution control investment completion ($x_{12}$) | + | 0.040 |

Some of the indicators need to be obtained through calculations, as described below. Energy intensity is measured by the ratio of energy consumption to energy output in each province; the coefficient of elasticity of energy production is measured by the ratio of the average annual growth rate of total energy production to the average annual growth rate of the economy in each province; and the R&D intensity is measured by the ratio of the amount of R&D investment to the Gross Regional Product in each province.

## 3.2 Data sources and descriptive statistics

The sample of this paper comes from 30 provinces or autonomous regions in China, and due to missing data, Tibet Autonomous Region, Hong Kong and Macao Special Administrative Regions, and Taiwan are not included in the sample this time. Considering the availability, the data in this paper mainly come from China Energy Statistical Yearbook (2004-2021), China Statistical Yearbook (2004-2021), CEEC Statistical Database (http://db.cei.cn), and China Carbon Accounting Database (https://www.ceads.net.cn). The data of some indicators (energy intensity; energy production elasticity coefficient; R&D intensity) are mainly derived by calculation. Considering the validity, this paper supplements a small amount of missing data for some provinces by linear interpolation method using the average annual growth.

**Tab.2 Descriptive statistics for variables**

| Variable | Obs. | Mean | Std. Dev. | Min | Max |
| --- | --- | --- | --- | --- | --- |
| $X_1$ | 540 | 0.5498313 | 0.3171487 | 0.1725252 | 1.750604 |
| $X_2$ | 540 | 0.0678977 | 7.161674 | -160.5337 | 10.10243 |
| $X_3$ | 540 | 7821079 | 6333499 | 273500 | 3.91e+07 |
| $X_4$ | 540 | 1.550029 | 1.109326 | 0.1802525 | 6.53 |
| $X_5$ | 540 | 106891.2 | 133965.6 | 1209 | 885248 |
| $X_6$ | 540 | 3270621 | 7620261 | 1885.29 | 7.01e+07 |
| $X_7$ | 540 | 3512.527 | 2171.85 | 221.7 | 14209.49 |
| $X_8$ | 540 | 119631 | 92268.7 | 4.56 | 434299.9 |
| $X_9$ | 540 | 97.57494 | 41.34241 | 0 | 228.09 |
| $X_{10}$ | 540 | 33.31398 | 17.94394 | 4 | 66.8 |
| $X_{11}$ | 540 | 3.1143676 | 3.739818 | 0.3791894 | 33.11152835 |
| $X_{12}$ | 540 | 19.058695 | 18.748823 | 0.0476 | 141.6 |

### 3.3 Methods

#### 3.3.1 Extreme value standardization

In order to eliminate the effect of the data magnitude, the data need to be normalized to the extremes[56]. The expression is:

$$y_{ij} = \begin{cases} \dfrac{max(x_j) - x_{ij}}{max(x_j) - min(x_j)} & - \\ \dfrac{x_{ij} - min(x_j)}{max(x_j) - min(x_j)} & + \end{cases} \quad (1)$$

In the formula, $y_{ij}$ denotes the raw indicator value of the $j$ th indicator of the $i$ th province, in which the inverse indicator needs to be transformed.

#### 3.3.2 Fixed-base efficacy coefficient method

In order to ensure that data on energy resilience indicators are comparable across years, this paper uses 2004 as the base period for the starting year of the sample and standardizes the raw data.

$$s_{ij}(t_k) = \begin{cases} \dfrac{max[x_j(t_1)] - x_{ij}(t_k)}{max[x_j(t_1)] - min[x_j(t_1)]} & - \\ \dfrac{x_{ij}(t_k) - min[x_j(t_1)]}{max[x_j(t_1)] - min[x_j(t_1)]} & + \end{cases} \quad (2)$$

where $x_{ij}(t_k)$ and $s_{ij}(t_k)$ denote the original and normalized data, respectively, and $max[x_j(t_1)]$ and $min[x_j(t_1)]$ denote the maximum and minimum values of the original data in the base period, respectively.

#### 3.3.3 linear weighting scheme

Combining the vector of weight coefficients sought and the normalized indicator values, the linear weighting method can be applied to obtain the energy resilience value of each province with the following expression:

$$Q_i(t_k) = \sum_{j=1}^{m} w_j s_{ij}(t_k) \quad (3)$$

#### 3.3.4 Centre of gravity-standard deviation ellipse

Lefever first proposed the definition of standard deviation ellipse in 1926 and argued that the concentration or dispersion of a system can be represented by the area of the standard deviation ellipse[57]. The standard deviation ellipse is a spatial statistical method that accurately reflects the spatial multifaceted characteristics of elements. It uses the main parameters such as mean center, azimuth, long and short axes to quantitatively reveal the spatial distribution pattern of elements, and is now used in many fields[58].

The formula is as follows:

$$\overline{X} = \frac{\sum_{i=1}^{n} W_i X_i}{\sum_{i=1}^{n} W_i}, \overline{Y} = \frac{\sum_{i=1}^{n} W_i Y_i}{\sum_{i=1}^{n} W_i} \quad (4)$$

$$S = \pi \sigma_x \sigma_y \quad (5)$$

Where $n$ denotes the number of provinces, $(X,Y)$ denotes the geographic coordinates of each province, $(\overline{X},\overline{Y})$ denotes the coordinates of the centre of gravity of energy resilience, and $\sigma_x$, $\sigma_y$ denote the standard deviation along the x-axis and y-axis, respectively.

### 3.3.5 kernel density estimate

Kernel density estimation belongs to one of the nonparametric estimation methods, which can get the information of distribution pattern according to the characteristics of the data itself, and overcome the error caused by specifying a certain distribution pattern in advance[59]. In this paper, the distributional dynamics of energy resilience in China is measured by unconditional kernel density estimation, static kernel density estimation and dynamic kernel density estimation.

The unconditional probability density expression is:

$$f(x) = \frac{1}{Nh} \sum_{i=1}^{N} K\left(\frac{X_i - x}{h}\right) \quad (6)$$

where $X_i$ denotes the observation, $x$ denotes the mean of the observations, $N$ denotes the number of sample observations, $K\left(\frac{X_i - x}{h}\right)$ denotes the Gaussian kernel function, and $h$ denotes the bandwidth.

The spatial static and spatial dynamic conditional probability density expressions are:

$$g(y|x) = \frac{f(x,y)}{f(x)} \quad (7)$$

$$f(x, y) = \frac{1}{Nh_x h_y} \sum_{i=1}^{N} K_x\left(\frac{X_i - x}{h_x}\right) K_y\left(\frac{Y_i - y}{h_y}\right) \quad (8)$$

Where $f(x, y)$ is the joint probability density of $x$ and $y$.

### 3.3.6 Geographical detector

Geo-detector is a statistical method for detecting spatial dissimilarity and revealing the driving forces behind it, including four detectors, namely factor detection, interaction detection, risk detection, and ecological detection, which argues that if the independent variables have an effect on the dependent variables, they will show similarity in their spatial distributions[60], and the specific formulas and measurements can be found in the literature[61].

# 4 Results

## 4.1 The Time-Series Evolution of China's Energy Resilience

Matlab is used to solve equation (1), and the weights of the specific indicators are listed in Table 1. The value of China's energy resilience and the national average value in the period 2004-2021 are obtained from the combination of equation (2) and equation (3), as shown in Table 2.

China's energy resilience increases steadily from 2004 to 2021, and the value of national average energy resilience increases from 0.36 in 2004 to 0.71 in 2021, with an average annual growth rate of 4.08%.The level of energy resilience of Chinese provinces also shows an increasing trend year by year, and the ten provinces with the highest average value of energy resilience from 2004 to 2021 are Guangdong, Shandong, Jiangsu, Zhejiang, Henan, Anhui, Hebei, Inner Mongolia, Shaanxi and Sichuan, with the energy resilience scores of Guangdong, Shandong, Jiangsu, Zhejiang, Henan, Hebei, Inner Mongolia, Shaanxi and Sichuan being higher than the national average value of energy resilience each year, and the ranking of The top-ranked Guangdong's energy resilience average value is 1.53, which is 7.65 times higher than that of the bottom-ranked Qinghai. Meanwhile, the top 10 provinces in terms of average annual growth rate of energy resilience from 2004 to 2021 are Ningxia, Anhui, Qinghai, Shandong, Guangdong, Jiangsu, Henan, Hebei, Hubei and Zhejiang. Among them, Ningxia has the highest average annual growth rate of energy resilience value in the country, reaching 9.53%, which is 3.27 times higher than that of Yunnan, which has the lowest average annual growth rate.

Overall, the level of national energy resilience in the period 2004-2010 was low, and the average value of national energy resilience did not exceed the average value of overall energy resilience in the period 2004-2021, which was 0.71. In 2013, the 12th Five-Year Plan for Energy Development was formally released, which mainly sets out China's guiding ideology, basic principles, development goals, key tasks, and the development of energy resources. principles, development goals, key tasks and policy measures, and put forward the control target of "total national energy consumption of 4 billion tons of standard coal" in 2015[62].The overall energy resilience of the whole country is significantly better in the period 2013-2019, indicating that in the "Energy Development The average annual growth rate of national energy resilience in the period 2019-2021 is only 2.51%, which may be due to the impact of the New Crown Epidemic Incident (NCEI), resulting in a slowdown in the growth rate of energy resilience.

**Tab.3 Energy Resilience Index by Province in China**

| Province | 2004 | 2007 | 2010 | 2013 | 2016 | 2019 | 2021 | Average |
|---|---|---|---|---|---|---|---|---|
| Beijing | 0.57 | 0.65 | 0.69 | 0.81 | 0.85 | 0.98 | 1.02 | 0.79 |
| Tianjin | 0.27 | 0.38 | 0.44 | 0.56 | 0.57 | 0.57 | 0.59 | 0.48 |
| Hebei | 0.39 | 0.51 | 0.69 | 1.00 | 1.11 | 1.24 | 1.27 | 0.89 |
| Shanxi | 0.39 | 0.59 | 0.74 | 0.95 | 0.95 | 0.69 | 0.65 | 0.71 |
| Inner Mongolia | 0.37 | 0.61 | 0.88 | 1.07 | 1.09 | 0.95 | 0.95 | 0.85 |
| Liaoning | 0.43 | 0.59 | 0.80 | 0.92 | 0.76 | 0.82 | 0.85 | 0.74 |
| Jilin | 0.31 | 0.39 | 0.51 | 0.52 | 0.56 | 0.54 | 0.54 | 0.48 |
| Heilongjiang | 0.39 | 0.51 | 0.62 | 0.66 | 0.56 | 0.60 | 0.65 | 0.57 |
| Shanghai | 0.34 | 0.46 | 0.52 | 0.57 | 0.66 | 0.71 | 0.82 | 0.58 |
| Jiangsu | 0.57 | 0.68 | 0.96 | 1.30 | 1.59 | 1.80 | 1.94 | 1.26 |
| Zhejiang | 0.57 | 0.74 | 0.88 | 1.15 | 1.33 | 1.63 | 1.73 | 1.15 |
| Anhui | 0.33 | 0.47 | 0.72 | 1.08 | 1.14 | 1.26 | 1.43 | 0.92 |
| Fujian | 0.40 | 0.48 | 0.61 | 0.80 | 0.87 | 0.93 | 1.06 | 0.74 |
| Jiangxi | 0.33 | 0.38 | 0.51 | 0.60 | 0.70 | 0.83 | 0.96 | 0.62 |
| Shandong | 0.60 | 0.79 | 1.08 | 1.44 | 1.73 | 1.88 | 2.12 | 1.38 |
| Henan | 0.42 | 0.57 | 0.77 | 0.87 | 1.13 | 1.31 | 1.41 | 0.93 |
| Hubei | 0.38 | 0.45 | 0.63 | 0.75 | 0.87 | 1.01 | 1.16 | 0.75 |
| Hunan | 0.39 | 0.50 | 0.67 | 0.85 | 0.88 | 0.96 | 1.12 | 0.77 |
| Guangdong | 0.68 | 0.88 | 1.21 | 1.63 | 1.77 | 2.21 | 2.37 | 1.53 |
| Guangxi | 0.33 | 0.40 | 0.53 | 0.65 | 0.73 | 0.81 | 0.96 | 0.63 |
| Hainan | 0.18 | 0.15 | 0.22 | 0.28 | 0.28 | 0.28 | 0.31 | 0.24 |
| Chongqing | 0.29 | 0.37 | 0.46 | 0.55 | 0.59 | 0.62 | 0.70 | 0.51 |
| Sichuan | 0.45 | 0.57 | 0.74 | 0.95 | 0.99 | 1.06 | 1.10 | 0.84 |
| Guizhou | 0.23 | 0.29 | 0.37 | 0.48 | 0.53 | 0.57 | 0.56 | 0.43 |
| Yunnan | 0.27 | 0.36 | 0.44 | 0.55 | 0.54 | 0.46 | 0.44 | 0.44 |
| Shanxi | 0.37 | 0.50 | 0.73 | 1.02 | 1.02 | 1.09 | 1.12 | 0.84 |
| Gansu | 0.14 | 0.22 | 0.35 | 0.49 | 0.44 | 0.34 | 0.37 | 0.34 |
| Qinghai | 0.07 | 0.11 | 0.15 | 0.23 | 0.27 | 0.27 | 0.30 | 0.20 |
| Ningxia | 0.07 | 0.14 | 0.24 | 0.29 | 0.36 | 0.32 | 0.34 | 0.25 |
| Xinjiang | 0.19 | 0.29 | 0.42 | 0.71 | 0.68 | 0.51 | 0.47 | 0.47 |
| National Average | 0.36 | 0.47 | 0.62 | 0.79 | 0.85 | 0.91 | 0.98 | 0.71 |

**Note: Due to space constraints, the measurements for 2004, 2007, 2010, 2013, 2016, 2019, and 2021 were selected.**

In order to reflect the spatial distribution characteristics of China's energy resilience more intuitively, this paper uses ArcGIS10.8 software to visualize the dynamic evolution of China's energy resilience, and adopts the natural discontinuity grading method to divide the energy resilience level of China's 30 provinces or autonomous regions into five types, namely, low-level zone, lower-level zone, medium-level zone, higher-level zone, and high-level zone, as shown in Figure 1.

Overall, China's energy resilience level shows a zigzagging development trend from 2004 to 2021. China's energy resilience high-level and higher-level zones are distributed in the central and eastern regions, while low-level and lower-level zones are distributed in the northwestern, southwestern and northeastern regions. The proportion of low-level zones, higher-level zones and high-level zones shows a downward and then upward trend, while the proportion of lower-level zones and medium-level zones shows a first upward and then downward trend, and the spatial imbalance of China's energy resilience is more obvious.

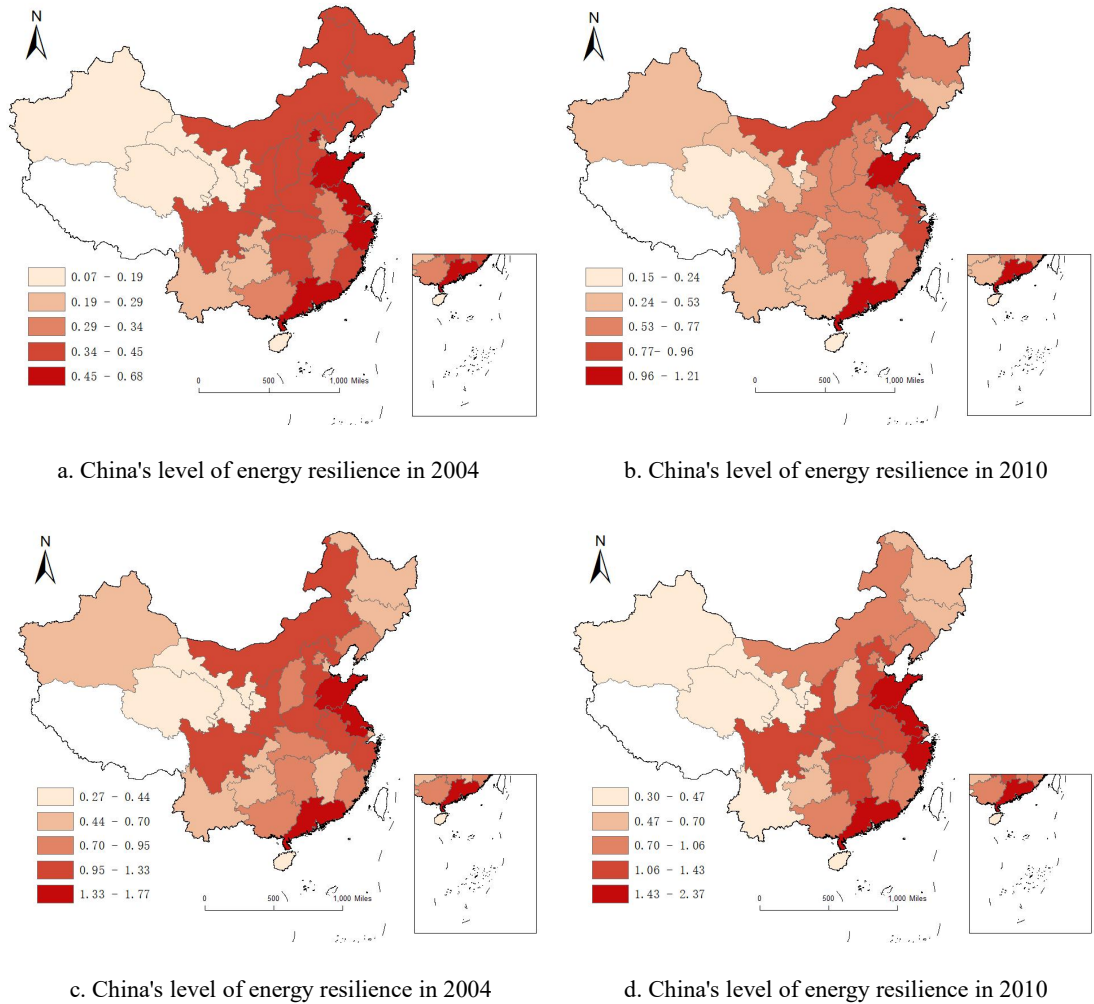

a. China's level of energy resilience in 2004

b. China's level of energy resilience in 2010

c. China's level of energy resilience in 2004

d. China's level of energy resilience in 2010

**Fig.1 Spatial distribution of energy resilience levels in China**

Specifically, Shandong Province and Guangdong Province have been in the high-level zone, Qinghai Province, Ningxia Hui Autonomous Region and Hainan Province have been in the low-level zone, and the energy resilience levels of the remaining provinces have changed to some extent. In 2004, China's energy resilience value range was [0.07,0.68], with higher-level zones predominating, accounting for 36.67 per cent of the total area, and low-level, lower-level, medium-level and high-level zones distributed more evenly. The distribution of low-level zone, lower-level zone, medium-level zone and high-level zone is relatively balanced, the proportion of higher-level zone and high-level zone is the highest in the period of 2004-2021, and the proportion of lower-level zone and medium-level zone is the lowest. 2010, China's energy resilience value range is [0.15,1.21], mainly medium-level zone, accounting for 36.67% of the total region, the proportion of the number of provinces of the lower level zone and medium-level zone is the highest in the period of 2004-2021, and the proportion of high level zone is the highest in the period of 2004-2021. In 2021, the number of provinces in the lower level zone and medium level zone was the highest, and the proportion of high level zone was the lowest in the period of 2004-2021, and only two provinces, Shandong Province and Guangdong Province, belonged to the high level zone. In 2016, China's energy resilience value range was [0.27,1.77], with the lower level zone dominating, accounting for 30% of the total region, and the increase or decrease in the number of provinces in the various level zones was small, and the level of the provinces basically remained the same or only one province's level was changed. In 2021, China's energy resilience value range will be [0.30,2.37], dominated by the medium-level zone and the higher-level zone, both accounting for 23.33 percent of the total region, with an increase in the number of provinces in the low-level zone and the higher-level zone, a decrease in the number of provinces in the lower-level zone and the medium-level zone, and the number of provinces in the high-level zone remaining unchanged.

## 4.2 The dynamic evolution of energy resilience in China

### 4.2.1 Centre of gravity-standard deviation elliptic analysis

In order to further study the dynamic evolution of China's energy resilience, this paper uses ArcGIS 10.8 software to draw the center of gravity-standard deviation ellipse dynamic evolution of China's energy resilience, as shown in Figure 2. The specific ellipse parameters are shown in Table 4.

**Tab.4 Parameters of China's energy resilience ellipse**

| Year | Key coordinates | Long semi-axis | Short semi-axis | Elliptical corner | Area |
| --- | --- | --- | --- | --- | --- |
| 2004 | (114.18,33.96) | 1135.78 | 833.36 | 21.86 | 2973560.24 |
| 2007 | (114.01,34.28) | 1119.27 | 866.08 | 22.27 | 3045388.80 |
| 2010 | (113.80,34.27) | 1108.79 | 883.37 | 21.01 | 3077101.48 |
| 2013 | (113.39,34.15) | 1075.14 | 941.87 | 18.20 | 3181309.02 |
| 2016 | (113.50,33.88) | 1040.45 | 918.88 | 15.79 | 3003515.56 |
| 2019 | (113.95,33.49) | 1035.82 | 850.86 | 15.90 | 2768804.37 |
| 2021 | (114.07,33.32) | 1029.76 | 827.77 | 15.82 | 2677907.51 |

From 2004 to 2021, the standard deviation ellipse of China's energy resilience showed a northeast-southwest pattern, covering most of the central and eastern regions, and shifted to the southwest as a whole, with the total area decreasing by 204755.87km². The center of gravity of China's energy resilience distribution has been in the territory of Henan Province, and the direction of the center of gravity shift is in the north-west-south-west-south-east direction, and the direction of the shift is in the south-east direction in 2013-2021, indicating that the level of China's energy resilience in the south-east direction has increased. The long half-axis of the standard deviation ellipse shows a continuous shortening trend from 1135.78 km in 2004 to 1029.76 km in 2021, indicating that China's energy resilience is clustered in the northeast-southwest direction. The short half-axis of the standard deviation ellipse shows a tendency of lengthening and then shortening, indicating that the evolution trend of China's energy resilience in the northwest-southeast direction is unstable for the time being. The turning angle of the standard deviation ellipse shows a trend of increasing and then decreasing, from 21.86° in 2004 to 22.27° in 2007, and then decreasing to 15.82° in 2021, indicating that China's energy resilience is shifting counterclockwise.

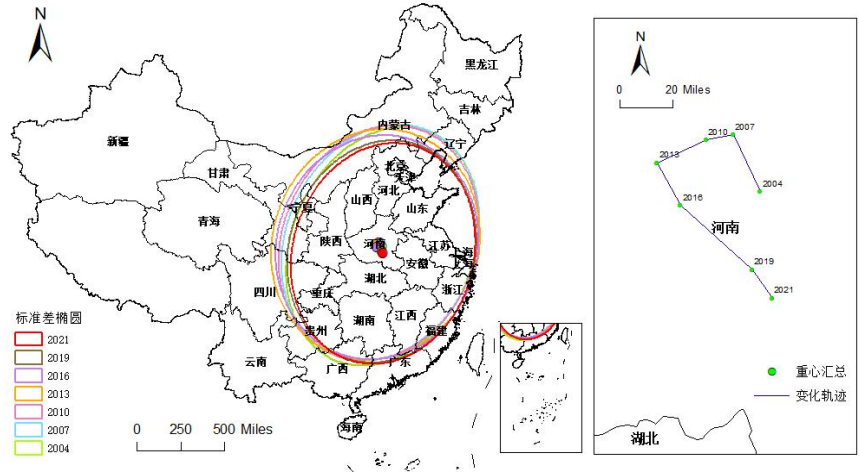

**Fig. 2 Standard deviation ellipse and center of gravity trajectory of China's energy resilience**

**4.2.2 Kernel density analysis**

In order to further reveal the dynamic evolution characteristics of China's energy resilience, this paper carries out kernel density estimation of China's energy resilience level from 2004 to 2021. First, the unconditional kernel density estimation method is used to examine the trend of China's energy resilience level from year t to year t+3; second, the static kernel density estimation method under spatial conditions is used to explore the spatial relationship between the energy level of each province and that of its neighboring provinces during the same period; finally, the dynamic kernel density estimation method under spatial conditions is used to reveal the impact of the energy resilience level of neighboring provinces in year t on the energy resilience level of the present province's energy resilience level in year t+3. Figures 3, 4 and 5 show the kernel densities and their contours of China's energy resilience under unconditional, spatial static and spatial dynamic conditions, respectively.

(1) Unconditional kernel density estimation

If the majority of the graph is centered on the positive 45° diagonal, it indicates that the level of energy resilience in year t+3 is the same as it was and has not changed; if the graph is rotated by 45° along the counterclockwise direction, it indicates that energy resilience is the same in all provinces in year t+3, and that there is a convergence in the growth of the provinces; in the case that if the 45° line is rotated by 90° along the counterclockwise direction, there is a sudden change, and the provinces with a high (low) level of energy resilience turn into the provinces with a low (high) level provinces[63]. As shown in Fig. 9b, the unconditional kernel density estimation probability body of China's energy resilience level is distributed along the positive 45° diagonal line, which indicates that the energy resilience level of each province is more persistent and less mobile, and that although the energy resilience level in year t+3 is improved compared to that of year t will be improved, the probability of a large change is low. In addition, the wave peak located near 0.15 on the x-axis is slightly higher than the 45° diagonal and parallel to y=0.25, and the wave peak located near 2.2 on the x-axis is slightly higher than the 45° diagonal and parallel to y=2.4, indicating that the provinces with energy resilience levels lower than 0.15 under the unconditional assumptions tend to be concentrated at 0.25 in year t+3, which is consistent with the above Hainan, Gansu, Qinghai and Ningxia provinces' The change characteristics of energy resilience level are consistent; the provinces with energy resilience level lower than 2.2 tend to concentrate at 2.4 in t+3 years, which is consistent with the change characteristics of energy resilience level of Guangdong province in 2019-2021 above.

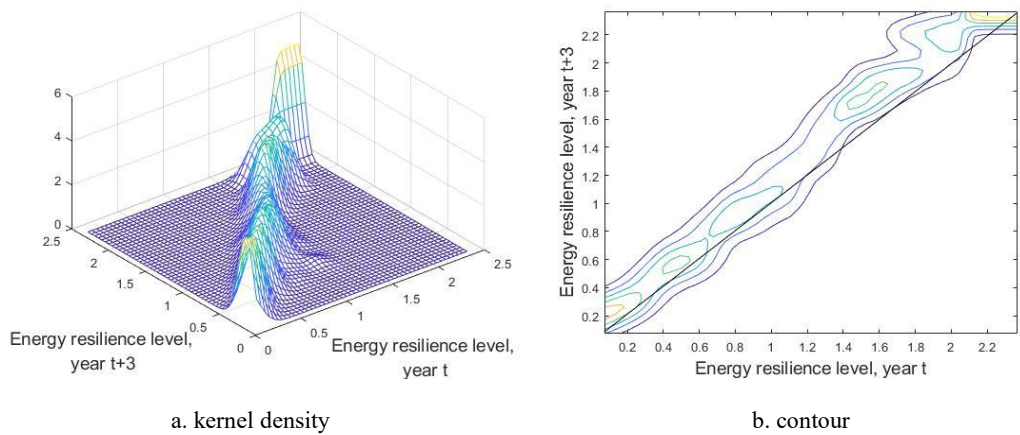

a. kernel density          b. contour

Fig.3 Unconditional kernel density of energy resilience in China and its contours

(2) Spatial static kernel density estimation

In the spatial conditional static kernel density estimation, if the probability body is mainly concentrated near the positive 45° diagonal, the energy resilience level of each province is in a convergence pattern with the neighboring provinces, in other words, there is a positive spatial correlation of the energy resilience level between the neighboring provinces, and the high-level provinces are clustered with the high-level provinces, and the low-level provinces are clustered with the low-level provinces[64]. As can be seen from Fig.10b, when the neighboring provinces have a low level of energy resilience, the main body of the probability is concentrated in the y-axis 0.2-0.6, indicating that the neighboring provinces with a low level of energy resilience do not play a large role in improving the level of energy resilience in the province. When the neighboring provinces energy resilience level in 1-1.4, the probability of the main distribution along the 45° diagonal, at this time between the neighboring provinces energy resilience level is positive spatial correlation, inter-provincial synergistic development can jointly improve the level of energy resilience. When the energy resilience of neighboring provinces is at 1.6-2.4, the probability body is parallel to the x-axis, indicating that when the energy resilience level of neighboring provinces is too high, the energy resilience level of this province will not be further improved due to the impact, and it is necessary to formulate an economic policy in line with the local situation, and to promote leapfrog improvement of energy resilience level through industrial upgrading and technological innovation.

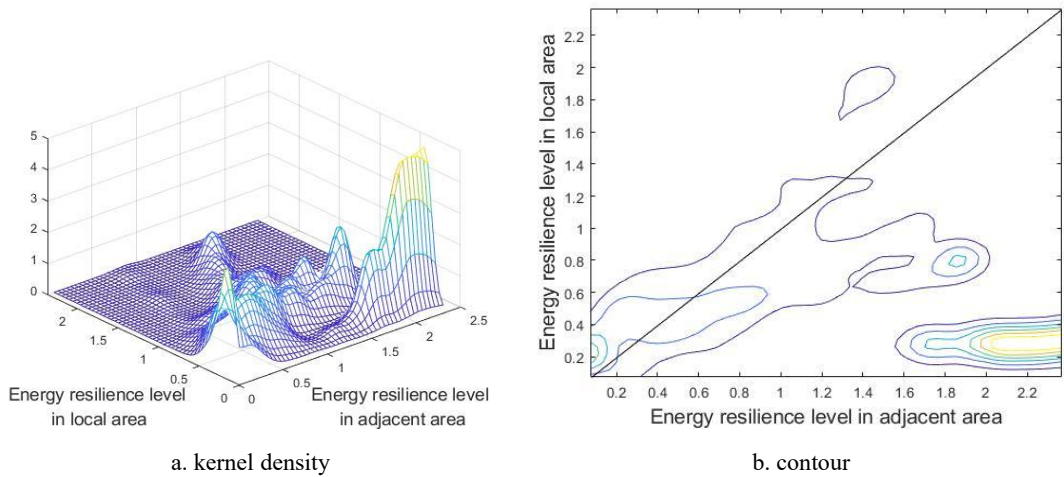

a. kernel density                                b. contour

Fig.4 Spatial static kernel densities and their contours for energy resilience in China

(3) Spatial dynamic kernel density estimation

This paper considers the time span on the basis of spatial static kernel density estimation to explore the impact of neighboring provinces' energy resilience level on the energy resilience level of this province in year t+3. The comparison of Fig.10b and Fig.11b shows that the impact of neighboring provinces' energy resilience levels on the home province under the spatial dynamic condition is almost the same as that under the spatial static condition, suggesting that the time condition does not play a significant role in driving the change of the home province's energy resilience level regardless of the energy resilience level of the neighboring province.

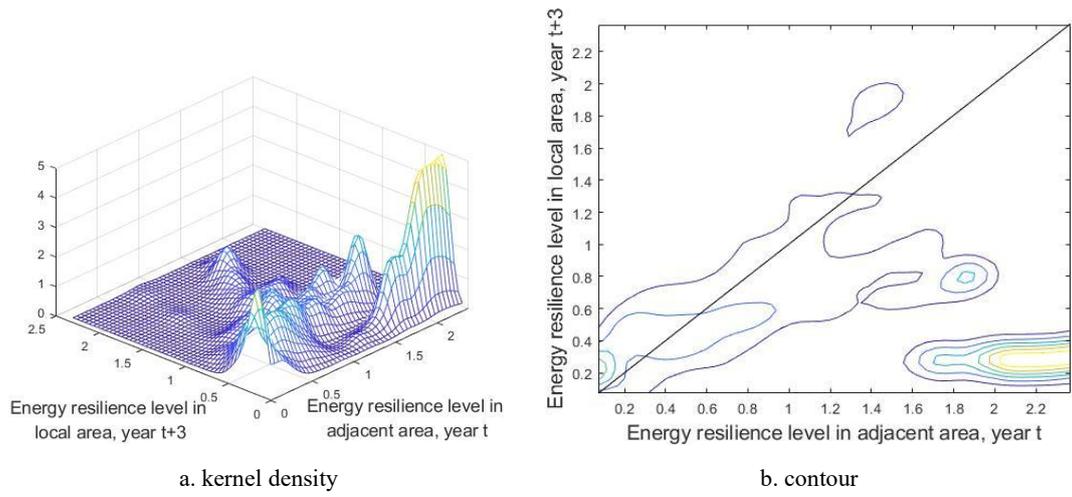

a. kernel density	b. contour

**Fig. 5 Spatial dynamic kernel density and its contours for energy resilience in China**

## 4.3 Factors influencing China's energy resilience

### 4.3.1 factor selection

In order to explore the influencing factors of China's energy resilience, this paper, on the basis of existing studies and considering the availability and operability of data, selects the factors that may affect China's energy resilience such as economy, population, employment, social security, and technological innovation[65,66,67,68], and uses geo-detectors to detect the separate economic, demographic, employment, education, social security, and technological innovation on China's energy resilience effects and interaction effects. The specific variables are as follows: Economy: GDP ($x_1$); Population: number of permanent residents ($x_2$); Employment: number of employees ($x_3$); Social security: number of employees' basic pension and medical insurance participants ($x_4$); Technological innovation: number of patent applications in high-tech industries ($x_5$).

The above data are obtained from the CEIC statistical database (http://db.cei.cn) and the China Economic and Social Big Data Research Platform (https://data.cnki.net/).

### 4.3.2 Geo-detector factor detection

In this paper, we use R-studio software to optimally discretize the data of the above five detection factors, transform them from continuous data to the category data needed by the geo-detectors, and then use the geo-detectors to detect the factors and interacting factors, and the magnitude of the influence ability is reflected by the Q-value, while the significance of the results is shown by the P-value, P*** < 0.01, P** < 0.01, specific detection results are shown in Table 5.

Tab.5 Detection results of impact factors of China's energy resilience

| Year | 2004 | | 2010 | | 2016 | | 2021 | |
|---|---|---|---|---|---|---|---|---|
| Variable | Q value | P value | Q value | P value | Q value | P value | Q value | P value |
| $X_1$ | 0.778 | 0.000*** | 0.722 | 0.000*** | 0.674 | 0.000*** | 0.755 | 0.000*** |
| $X_2$ | 0.411 | 0.038** | 0.561 | 0.004*** | 0.616 | 0.001*** | 0.627 | 0.001*** |
| $X_3$ | 0.542 | 0.005*** | 0.545 | 0.005*** | 0.622 | 0.001*** | 0.689 | 0.000*** |
| $X_4$ | 0.714 | 0.000*** | 0.666 | 0.000*** | 0.659 | 0.000*** | 0.743 | 0.000*** |
| $X_5$ | 0.625 | 0.001*** | 0.426 | 0.034** | 0.533 | 0.006*** | 0.754 | 0.000*** |

Through comparative observation, the mean values of the five variables at the four time points are 0.732, 0.554, 0.600, 0.696 and 0.585, respectively. $x_1$ (GDP) has the highest mean value, indicating that it has the greatest impact on China's energy resilience, and $x_2$ (number of resident population) has the smallest mean value, indicating that it has the least impact on China's energy resilience. Through time observation, the largest influence factor of China's energy resilience in 2004, 2010, 2016, and 2021 is $x_1$ (GDP), with Q values of 0.778, 0.722, 0.674, and 0.755, respectively.

Firstly, the economic level is an important determinant of China's energy resilience, and to improve China's energy resilience, it is necessary to formulate corresponding strategies according to the level of development and development advantages of different regions, to expand domestic demand, to improve productivity, to promote coordinated regional development, and to enhance economic strength. Secondly, the mean values of $x_4$ (the number of employees' basic pension and medical insurance participants) and $x_3$ (the number of employees) are 0.696 and 0.600, respectively, ranking the second and the third, and $x_4$ was the second most important influence factor of China's energy resilience in 2004, 2010, and 2016, which indicates that social security and employment can influence China's energy resilience to a certain degree, mainly due to the following reasons: first, social security and Firstly, social security and employment can provide human resource reserves for the energy industry, and secondly, social security and employment can indirectly affect the economic level while increasing the consumption level of residents. Finally, over time, the ability of technological innovation to influence China's energy resilience gradually improves, and $x_5$ (the number of patent applications in high-tech industries) becomes the second largest influence factor of China's energy resilience in 2021, with a Q value of 0.754. China's energy resilience relies to some extent on the level of technological innovation, which is mainly due to the fact that technological advances can improve the efficiency of energy utilization and reduce the cost of energy and pollution emissions. costs, reduce pollution emissions, etc.

**Tab.6 Detecting the interactions of energy resilience impact factors in China**

| 2004 | | 2010 | | 2016 | | 2021 | |
|---|---|---|---|---|---|---|---|
| interaction | Q value | interaction | Q value | interaction | Q value | interaction | Q value |
| $X_2 \cap X_4$ | 0.904 | $x_2 \cap x_5$ | 0.921 | $x_1 \cap x_5$ | 0.891 | $x_2 \cap x_5$ | 0.917 |
| $X_1 \cap X_2$ | 0.893 | $x_2 \cap x_4$ | 0.902 | $x_2 \cap x_4$ | 0.842 | $x_1 \cap x_5$ | 0.907 |
| $X_1 \cap X_3$ | 0.888 | $x_1 \cap x_5$ | 0.882 | $x_1 \cap x_4$ | 0.838 | $x_3 \cap x_5$ | 0.904 |
| $X_3 \cap X_4$ | 0.882 | $x_3 \cap x_4$ | 0.873 | $x_4 \cap x_5$ | 0.831 | $x_4 \cap x_5$ | 0.874 |
| $X_3 \cap X_5$ | 0.856 | $x_3 \cap x_5$ | 0.872 | $x_3 \cap x_4$ | 0.829 | $x_1 \cap x_2$ | 0.808 |
| $X_1 \cap X_5$ | 0.829 | $x_1 \cap x_4$ | 0.826 | $x_1 \cap x_3$ | 0.802 | $x_1 \cap x_4$ | 0.801 |
| $X_1 \cap X_4$ | 0.813 | $x_1 \cap x_3$ | 0.766 | $x_1 \cap x_2$ | 0.773 | $x_3 \cap x_4$ | 0.796 |
| $X_4 \cap X_5$ | 0.811 | $x_1 \cap x_2$ | 0.752 | $x_2 \cap x_5$ | 0.764 | $x_1 \cap x_3$ | 0.795 |
| $X_2 \cap X_5$ | 0.805 | $x_4 \cap x_5$ | 0.700 | $x_3 \cap x_5$ | 0.764 | $x_2 \cap x_4$ | 0.795 |
| $X_2 \cap X_3$ | 0.665 | $x_2 \cap x_3$ | 0.550 | $x_2 \cap x_3$ | 0.637 | $x_2 \cap x_3$ | 0.683 |

To further investigate the influence of interaction factors on China's energy resilience, geo-detectors were used to detect interaction factors, and the results are shown in Table 6. The results of the interaction factor detection show that, first, the explanatory power of most of the interaction factors is higher than the highest influence of the single-factor combination, and the interaction is of the enhanced type. Meanwhile, the mean values of the explanatory power of all interaction factors are greater than the mean value of the influence of a single factor, indicating that the two-factor interaction increases the explanatory power of China's energy resilience. Second, nine, six, six and six two-factor interactions have an explanatory power of more than 0.8 in 2004, 2010, 2016 and 2021, respectively, with a close relationship between the factors, and two-factor interactions have a significant impact on China's energy resilience in most years. Third, the explanatory power of some two-factor interactions reaches more than 0.9, such as $x_2$ (number of permanent residents) and $x_4$ (number of employees' basic pension and medical insurance participants) in 2004 and 2010, $x_2$ (number of permanent residents) and $x_5$ (number of patent applications in high-tech industries) in 2010 and 2021, $x_1$ (GDP) and $x_5$ (number of patent applications in high-tech industries) in 2021 , and $x_3$ (number of employees) and $x_5$ (number of patent applications in high-tech industries), which shows that technological innovation, jointly with other factors, has a more significant impact on China's energy resilience. In summary, China's energy resilience level is affected by the interaction of multiple factors, and in the future energy governance process, it should be viewed from a diversified perspective, cross-integrate the above influencing factors, and formulate combined policies.

# 5 Conclusion and discussion

On the basis of constructing an energy resilience evaluation system, this paper applies the combined dynamic evaluation method to measure the level of China's energy resilience from 2004 to 2021, analyses the spatio-temporal dynamic evolution of China's energy resilience through the center of gravity-standard deviation ellipse and kernel density estimation, and employs a geographic detector to detect the main influencing factors and interactions of China's energy resilience. The main conclusions are as follows:

First, in general, China's energy resilience level shows a zigzagging forward development trend from 2004 to 2021. From the time dimension, the overall national energy resilience level was low in 2004-2010, the overall national energy resilience was significantly better in 2013-2019, and the national energy resilience growth rate slowed down in 2019-2021. From the spatial dimension, China's energy resilience high-level and higher-level zones are distributed in the central and eastern regions, and low-level and lower-level zones are distributed in the northwest, southwest and northeast regions. The proportion of low-level zones, higher level zones and high level zones shows a downward and then upward trend, and the proportion of lower level zones and medium level zones shows a upward and then downward trend, and the spatial imbalance of China's energy resilience is more obvious.

Second, from 2004 to 2021, the spatial dynamics of China's energy resilience level evolved into a northeast-southwest direction, and the whole moved to the southwest, with the total coverage area decreasing by 204755.87km². The center of gravity of China's energy resilience level distribution has been in the territory of Henan Province, and the direction of the center of gravity shift is in the north-west-south-west-south-east direction, which is improved in the south-east direction, and the evolution trend in the north-west-south-east direction is unstable for the time being, and the overall trend is counter-clockwise and constantly shifting.

Thirdly, when the energy resilience level of neighboring provinces is low, it has little effect on the improvement of the energy resilience level of the province, when the energy resilience level of neighboring provinces is 1-1.4, it has a positive spatial correlation with the energy resilience level of the province, and when the energy resilience level of neighboring provinces is too high, the energy resilience level of the province will not be further improved due to the influence. Other than that, time horizon does not play a significant role in driving energy resilience levels between neighboring provinces.

Fourth, GDP, the number of employees, the number of employees' basic pension and medical insurance participants and the number of patent applications in high-tech industries have a more significant impact on China's energy resilience, i.e., factors such as the economy, employment, social security and technological innovation will have a more significant impact on China's energy resilience level. At the same time, China's energy resilience is affected by the interaction of multiple factors, and technological innovation factors have a more significant impact on China's energy resilience level than other factors.

Based on the above conclusions, more attention should be paid to the spatial imbalance of energy resilience in future development. First, most of the high energy resilience zones are in the eastern region of China, while most of the low energy resilience zones are in the western region. The eastern region is economically developed, technologically advanced and has a favorable geographical location and perfect transportation facilities. Although energy is scarce, it has always been at the forefront of energy technology in recent years thanks to the west-to-east electricity transmission and west-to-east gas transmission strategies and the impact of renewable energy development plans. While the western region has inherent advantages such as high energy density and abundant resources, the complex terrain, relatively cumbersome transportation and harsh climate have led to a rather backward economy, which is not conducive to improving energy resilience. Therefore, it is necessary to narrow the gap between the eastern and western regions by developing differentiated regional development strategies to jointly promote China's high-quality energy development. Second, when the energy resilience level of neighboring provinces is 1-1.4, there is a positive spatial correlation with the energy resilience level of the province, and the synergistic development of neighboring provinces in the higher-level region is conducive to the joint improvement of the energy resilience level. Therefore, it is necessary to accelerate the transformation of the development model of the middle reaches of the Yangtze River city cluster and the Chengdu-Chongqing city cluster, create special industries for city clusters, and raise the regional energy resilience level to a higher level. Finally, China's energy resilience is mainly influenced by the joint effect of economic level and technological innovation, and in the future development, it should further improve the modern economic system, improve the vitality of the main market, and accelerate the formation of a higher level of open economic system, and at the same time, on the basis of technological innovation, establish a more perfect renewable energy technology system and promote the digital and intelligent modernization of the energy industry.

Compared with previous studies, this paper takes into account the dynamic evolution of China's energy resilience, explores in depth the evolution characteristics and development trend of China's energy resilience in different spaces and times, and determines the influence of influencing factors and the influence of the interaction of factors, but the spatial differences in energy resilience and its sources in different provinces are not analyzed in depth and need to be further explored in future studies.